\documentclass[11pt]{article}
\usepackage{amsmath,amssymb}
\usepackage{graphicx}
\numberwithin{equation}{section}
\def\beq#1 #2\e{\begin{equation}\label{#1}#2\end{equation}}
\def\bea#1 #2\e{\begin{align}\label{#1}#2\end{align}}
\def\bml#1 #2\e{\begin{multline}\label{#1}#2\end{multline}}
\def\bmlg#1\e{\begin{multline*}#1\end{multline*}}
\def\bgr#1 #2\e{\begin{gather}\label{#1}#2\end{gather}}
\let\bal\aligned \let\eal\endaligned
\let\bga\gathered \let\ega\endgathered
\def\bca{\begin{cases}}\def\eca{\end{cases}}
\def\lb#1 {\label{#1}}
\let\nn\nonumber
\let\er\eqref
\let\a\alpha
\def\am{\mathop{\rm am}}
\def\cG{\mathcal G}
\def\cH{\mathcal H}
\def\Corr{\mathop{\rm Corr}}
\def\Cov{\mathop{\rm Cov}}
\def\cP{\mathcal P}
\def\ct{{\rm const}}
\let\d\delta
\let\D\Delta
\let\dss\displaystyle
\let\dsl\displaylines
\def\dwa#1{\textstyle{\substack{#1}}}
\def\gd#1,#2,#3,{#1^{#2}{}_{#3}}
\def\gG{\mathfrak G}
\let\iy\infty
\let\k\kappa
\let\La\Lambda
\let\la\lambda
\def\m{{\mu\nu}}
\let\mt\mapsto
\def\mid{\kern1.5pt|\kern1.5pt}
\let\o\omega
\let\ov\overline
\def\pd#1 #2 {\frac{\partial #1}{\partial #2}}
\def\PD#1 #2 {\frac{d #1}{d #2}}
\def\Pd#1 #2 {\frac{\d #1}{\d #2}}
\let\q\quad
\def\R{\mathbb R}
\def\Re{\mathop{\rm Re}}
\let\s\sigma
\def\sn{\mathop{\rm sn}\nolimits}
\let\t\times
\let\th\theta
\def\Tr{\mathop{\rm Tr}\nolimits}
\def\Var{\mathop{\rm Var}\nolimits}
\let\vf\varphi
\let\wh\widehat
\let\wt\widetilde
\def\<#1>{\bigl\langle#1\bigr\rangle}
\def\bg(#1){\bigl(#1\bigr)}
\def\Bg(#1){\Bigl(#1\Bigr)}
\def\bgg(#1){\biggl(#1\biggr)}

\def\ch{characteristic}
\def\cn{constant}
\def\cd{coordinate}
\def\eq{equation}
\def\f{function}
\def\Hf{Hamilton function}
\def\ho{harmonic oscillator}
\def\hg{homogeneous}
\def\mc{mechanical}
\def\op{operator}
\def\qm{quantum}
\def\qz{quantiz}
\def\so{solution}
\def\sp{stochastic process}
\def\st{such that }
\def\tf{transformation}
\def\va{variable}
\def\wrt{with respect to }
\def\up#1{\uppercase{#1}}
\def\U{\expandafter\up}
\let\ti\textit
\let\TM\texttrademark

\author{M. W. Kalinowski\\
Pracownia Bioinformatyki, Instytut Medycyny Do\'swiadczalnej
i Klinicznej PAN,\\Pawi\'nskiego 5, 02-106 Warszawa\\
e-mail: markwkal1@gmail.com}
\title{Time as a stochastic process}

\begin{document}
\maketitle

\begin{abstract}
In the paper we consider an interesting possibility of a time as a stochastic
process in quantum mechanics. In order to do it we reconsider time as
a~mechanical quantity in classical mechanics and afterwards we quantize it.
We consider continuous and discrete time.
\end{abstract}

\vskip12pt
\rightline{\small
\vbox{\parindent0pt \leftskip0pt plus 1fill \rightskip0pt
Motto: \it The notion of time is not sufficiently well defined\\
to answer a problem of quantum time\\
\rm Gerard 't Hooft}}

\section{Introduction}
In this paper we consider a time as a \sp\ in quantum
mechanics. Time is considered as a parameter of an evolution of a mechanical
system (e.g.\ in a Schr\"odinger \eq).
It seems for us that it is not enough. We consider a time as a
mechanical quantity (Section~2), afterwards we quantize this quantity using
canonical quantization procedure (Section~3) and in Section~4 we consider
time as a \sp. It seems that our approach is a new one and has nothing to do
with any approaches to quantization of time. We give several examples of
mechanical systems with their own times. We give an example with a discrete
time.

\section{Time as a \mc\ quantity}
Let us define the following \f\ of a \mc\ system
\[
T=T(q_i,p_i,t), \q i=1,2,\dots,n,
\]
where $q_i$ are \cd s of the \mc\ system and $p_i$ are momenta conjugated to
those \cd s. $t$~is the usual time which plays here a r\^ole of a parameter
(see Ref.~\cite1).

Let
\beq2.1
\frac{dT}{dt}=c=\ct
\e
is an \eq\ of motion for such a \f.

It is easy to see that we can reparametrize $t$ to get the following \eq
\beq2.2
\frac{dT}{dt}=1.
\e
We neglect also a translation along $t$.

One gets
\beq2.3
\frac{dT}{dt}=\pd T t + \sum_{i=1}^n \Bigl(\pd T q_i \,\dot q_i +
\pd T p_i \,\dot p_i\Bigr)
= \pd T t +\sum_{i=1}^n \Bigl(\pd T q_i \,\pd H p_i - \pd T p_i \,
\pd H q_i \Bigr)= \pd T t + \{T,H\}
\e
where $H=H(p_i,q_i)$ is a Hamilton \f\ for the system and ``$\dot{\kern5pt }$''
means the differentiation \wrt $t$. We use of course the Hamilton \eq s
\beq2.4
\dot q_i=\pd H p_i \,, \q \dot p_i=-\pd H q_i \,,
\e
$\{\cdot,\cdot\}$ means the Poisson bracket.

Thus we get
\beq2.5
\pd T t +\{T,H\}=1.
\e
Eq.\ \er{2.5} is an \eq\ of motion for a \f\ which can be considered as a time
of the \mc\ system. Eq.~\er{2.5} can be solved giving us $T(q_i,p_i,t)$. Let
us consider a simple example of a Hamilton \f
\bgr2.6
H=\frac12 \sum_{i=1}^n \Bigl(\frac{p_i^2}{m_i} +K_i q_i^2\Bigr)\\
\o_i^2=\frac{K_i}{m_i}\,, \q \o_i=2\pi \nu_i=\frac{2\pi}{T_i}\,,
 \label{2.7}
\e
which is an $n$-dimensional harmonic oscilator \f. One gets
\beq2.8
\pd T t +\sum_{i=1}^n \Bigl(\pd T q_i \, \frac{p_i}{m_i}
-\pd T p_i K_iq_i\Bigr)=1.
\e
Equation \er{2.8} can easily be solved in terms of one arbitrary \f\ of
$2n$~variables
\beq2.9
\wt F=\wt F(x_1,y_1,x_2,y_2,\dots,x_n,y_n).
\e
One gets
\bml2.10
T(q_1,q_2,\dots,q_n,p_1,p_2,\dots,p_n,t)=T(q_i,p_i,t)\\
{}=t+\wt F\Bigl(\bigl(p_1\cos\o_1t+q_1\sqrt{K_1m_1}\,\sin\o_1t\bigr),
\Bigl(-q_1\cos\o_1t+p_1\,\frac{\sin\o_1t}{\sqrt{K_1m_1}}\Bigr), \kern60pt \\
\bigl(p_2\cos\o_2t+q_2\sqrt{K_2m_2}\,\sin\o_2t\bigr),
\Bigl(-q_2\cos\o_2t+p_2\,\frac{\sin\o_2t}{\sqrt{K_2m_2}}\Bigr),\\
\dots, \bigl(p_n\cos\o_nt+q_n\sqrt{K_nm_n}\,\sin\o_nt\bigr),
\Bigl(-q_n\cos\o_nt+p_n\,\frac{\sin\o_nt}{\sqrt{K_nm_n}}\Bigr)\Bigr).
\e
One can easily check \er{2.10} by simple calculations putting \er{2.10} into
Eq.~\er{2.8}.

The Cauchy initial value problem for Eq.~\er{2.8} can be defined by
\beq2.11a
T(q_i,p_i,0)=\wt F(p_1,-q_1,p_2,-q_2,\dots,p_n,-q_n), \q i=1,2,\dots,n.
\e
One can easily prove
\beq2.12a
T\Bigl(q_i(t_i),p_i(t_i),t+\sum_{j=1}^n l_jT_j\Bigr)
=T(q_i(t_i),p_i(t_i),t)+\sum_{j=1}^n l_jT_j,
\e
$l_j$ are integers and $t_i=t+\sum_{\substack{j=1\\j\ne i}}^nl_jT_j$.

If we disregard the dependence of~$t$ in $T(q,p,t)$, i.e.\ we consider
$T=T(q,p)$, we get
\beq2.11
\{T,H\}=1.
\e
In the one-dimensional case for a harmonic Hamilton \f
\beq2.12
H=\frac12\Bigl(\frac{p^2}{m} + Kq^2\Bigr)
\e
one gets two \so s
\beq2.13
T(q,p)=\bca
\dss -\frac1\o \arctan \Bigl(\frac p{\sqrt{Km}\,q}\Bigr)
+f\Bigl(\frac{p^2+Kmq^2}{2Km}\Bigr)\\
\dss \frac1\o \arctan \Bigl(\frac {\sqrt{Km}\,q}p\Bigr)
+f\Bigl(\frac{p^2+Kmq^2}{2Km}\Bigr)\eca
\e
where $f$ is an arbitrary \f\ of one variable and of $C^{(1)}$ class. One can
easily prove that $T(p(t),q(t))$ (where $p(t)$ and $q(t)$ are \so s of a
harmonic oscillator \eq\ of motion) is a periodic \f\ of~$t$. We can use
polar \cd s in the following way:
$$
y=\frac p{\sqrt{Km}}, \q x=q, \q x^2+y^2=r^2, \q \tan \vf=\frac yx,
$$
getting $\frac{p^2+Kmq^2}{Km}=r^2$, $\tan\vf=\frac p{\sqrt{Km}\,q}$ and
$$
T(q,p)=T(r,\vf)=\bca
\dss -\frac {\vf+k_i} \o + f\Bigl(\frac{r^2}2\Bigr)\\
\dss \frac {(\frac\pi2-\vf-k_i)}\o + f\Bigl(\frac{r^2}2\Bigr),\eca
\q i=1,2,3,4,
$$
for $-\frac\pi2< \vf<\frac\pi2$. Moreover, for polar \cd\ $0< \vf\le 2\pi$
and we should add some contants to normalize $\vf$ in four quadrants, i.e.\
$k_1=0$, $k_2=k_3=\pi$, $k_4=2\pi$.

We can
consider also a free motion of the \mc\ system, i.e.\ considering a Hamilton
\f
\beq2.14
H=\frac12 \sum_{i=1}^n \frac{p_i^2}m\,
\e
and the \eq
\beq2.15
\pd T t + \sum_{i=1}^n \pd T q_i \,\frac{p_i}{m_i}=1.
\e
Eq.\ \er{2.15} can be also solved and we get
\bml2.16
T(q_1,q_2,\dots,q_n,p_1,p_2,\dots,p_n,t)\\
{}= t+\wt F\Bigl(p_1,\Bigl(-q_1+\frac{p_1t}{m_1}\Bigr),
p_2,\Bigl(-q_2+\frac{p_2t}{m_2}\Bigr),\dots,
p_n,\Bigl(-q_n+\frac{p_nt}{m_n}\Bigr)\Bigr)
\e
where $\wt F(x_1,y_1,x_2,y_2,\dots,x_n,y_n)$ is a \f\ of $2n$ \va s of $C^{(1)}$
class.

In the case of Eq.\ \er{2.11} for a \Hf
\beq2.17
H=\frac12\,\frac{p^2}m
\e
one gets
\beq2.18
\pd T q \,\frac pm=1.
\e
Eq.\ \er{2.18} can be easily solved
\beq2.19
T =\Bigl(\frac mp\Bigr)q+f(p),
\e
where $f$ is an arbitrary \f\ of one \va\ and of $C^{(1)}$ class.

Let us define the following quantities:
\bea2.20
a_j&=\sqrt{\frac{m_j\o_j}{2\hbar}}\,\Bigl(q_j+\frac i{m\o}\,p_j\Bigr),\\
a_j^*&=\sqrt{\frac{m_j\o_j}{2\hbar}}\,\Bigl(q_j-\frac i{m\o}\,p_j\Bigr),
\lb2.21
\e
$j=1,2,\dots,n$. Using relations $\sqrt{K_j m_j}=\o_jm_j$ one gets
\bml2.22
T(t,q_j,p_j)=\wt T(t,a_j,a_j^*)=
t+\wt F\Bigl(-\sqrt{2m_1\o_1\hbar}\Re\bigl(ia_1e^{i\o_1t}\bigr),
-\sqrt{\frac{2\hbar}{m_1\o_1}}\Re\bigl(a_1e^{i\o_1t}\bigr),\\
-\sqrt{2m_2\o_2\hbar}\Re\bigl(ia_2e^{i\o_2t}\bigr),
-\sqrt{\frac{2\hbar}{m_2\o_2}}\Re\bigl(a_2e^{i\o_2t}\bigr),\\
\dots,
-\sqrt{2m_n\o_n\hbar}\Re\bigl(ia_ne^{i\o_nt}\bigr),
-\sqrt{\frac{2\hbar}{m_n\o_n}}\Re\bigl(a_ne^{i\o_nt}\bigr)\Bigr).
\e
A \cn\ $\hbar$ is here arbitrary and is not the Planck \cn.

The inverse \tf\ is as follows:
\bea2.23
q_j&=\sqrt{\frac{\hbar}{2m_j\o_j}}\,(a_j+a_j^*),\\
p_j&=-\sqrt{\frac{m_j\o_j\hbar}{2}}\,(a_j^*-a_j). \lb2.24
\e

Let us consider a different Hamiltonian
\beq2.25
H=\frac1{2m} \bigl(p_r^2+r^2p_\th^2+r^2\sin^2\th p^2_\vf\bigr)+V(r)
\e
where $V(r)=\frac kr$. In this way we take under consideration a Coulomb or a
Newton problem for two bodies if we identify $m$ as a reduced mass. From
Eq.~\er{2.3} one easily gets
\bml2.26
\pd T t + \biggl(\pd T r \cdot \frac{p_r}m - \pd T p_r \Bigl(
\frac rm\bigl(p_\th^2+\sin^2\th p_\vf^2\bigr)-\frac k{r^2}\Bigr)\biggr)\\
{}+\Bigl(\pd T \th{} \cdot \frac{r^2 p_\th}{m} - \pd T p_\th{} \cdot
\frac{r^2p_\vf^2\sin\th\cos\th}m \Bigr) +
\Bigl(\pd T \vf{} \cdot \frac{r^2\sin^2\th}m \,p_\vf\Bigr)=1.
\e
Eq.~\er{2.26} is a non\hg\ \eq. We consider also a \hg\ \eq\ (i.e.\ with zero
on the right hand side)
\[
T=T(t,r,p_r,\th,p_\th,\vf,p_\vf),
\]
$r,\th,\vf$ are usual spherical \cd s and $p_r,p_\th,p_\vf$ are moments
conjugated to them.

Eq.\ \er{2.26} is a linear PDE of the first order and can be solved by using
a characteristic method. One writes a system of \eq s for a characteristic
line. In this way we first consider \hg\ \eq\ and afterwards we find a
special \so\ to non\hg\ \eq s, i.e.\ Eq.\ \er{2.26}.
\refstepcounter{equation} \label{2.27} \def\nrw#1{\hfill (\theequation.#1)}
\def\li{\hskip100pt}
$$\dsl{
\li \PD t \tau{} =1 \nrw1 \cr
\li \PD r \tau{} =\frac{p_r}{m} \nrw2 \cr
\li \PD p_r \tau{} = -\Bigl( \frac rm \bigl(p_\th + \sin^2\th p_\vf^2\bigr)
-\frac k{r^2}\Bigr) \nrw3 \cr
\li \PD \th{} \tau{} = \frac{r^2p_\th}{m} \nrw4 \cr
\li \PD p_{\th} \tau{} = -\frac{r^2p_\vf^2\sin\th \cos\th}m \nrw5 \cr
\li \PD \vf{} \tau{} = \frac{r^2\sin^2\th}m \,p_\vf \nrw6 \cr
\li \PD p_{\vf} \tau{} = 0. \nrw7 \cr}
$$
$\tau$ is a natural parameter of the line. It is easy to see that
$p_\vf=c=\ct$.

We take $t=\tau$ and afterwards reparametrize the \eq s to a more convenient
parameter~$r$. Eventually one gets
\bea 2.28
t(r)&=\frac m{p_0}\cdot \root 3 \of {\frac{2k}d}\,\biggl(G_1\biggl(r,
\root3\of{\frac{d^2k^2}2}\biggr)-f\biggr) \\
r&=r \lb 2.29 \\
p_r(r)&=p_0 \exp\Bigl(-\frac kr-\frac{r^2d^2}2\Bigr) \lb2.30 \\
\th(r)&=\am\biggl(\frac{2k}{p_0}\,G_2\biggl(r,\root3\of{\frac{d^2k^2}w}\biggr)
+e \Bigm| \frac{c^2}{d^2}\biggr) \lb2.31 \\
p_\th(r)&=\sqrt{d^2-c^2\sn^2\biggl(\frac{2k}{p_0}\,G_2\biggl(r,\root3\of
{\frac{d^2k^2}2}\biggr) +e \Bigm|\frac{c^2}{d^2}}\biggr) \lb 2.32 \\
\vf(r)&=\frac dc \biggl (\frac{2k}{p_0}\,G_2\biggl(r,\root3\of
{\frac{d^2k^2}2}\biggr) +e - E\biggl(\am\biggl(\frac{2k}{p_0}\,G_2\biggl(r,\root3\of
{\frac{d^2k^2}2}\biggr) +e\Bigm| \frac{c^2}{d^2}\biggr)\Bigm| \frac{c^2}{d^2}
\biggr)\biggr)+\vf_0 \lb2.33 \\
p_\vf(r)&=c \lb2.34
\e
where $p_0,d,f,e,c,\vf_0$ are \cn s of integration,
\bea 2.35
G_1(r,a)&=\int_{1/2}^r dx\exp\Bigl(a\Bigl(\frac1x+x^2\Bigr)\Bigr)\\
G_2(r,a)&=\int_{1/2}^r dx\,x^2\exp\Bigl(a\Bigl(\frac1x+x^2\Bigr)\Bigr),\lb2.36
\e
$\am(x\mid k^2)$ is the Jacobi amplitude \f, i.e.\ the inverse \f\ of the
elliptic integral of the first order
\beq2.37
F(x\mid k^2)=\int_0^x \frac{d\th}{\sqrt{1-k^2\sin^2\th}}\,,
\e
$E(x\mid k^2)$ is the elliptic integral of the second order
\bgr2.38
E(x\mid k^2)=\int_0^x d\th\,\sqrt{1-k^2\sin^2\th},\\
\sn(\a\mid k^2)=\sin(\am(\a\mid k^2)) \lb2.39
\e
is the Jacobi elliptic \f.

Let us notice the following fact. The \f\ $G_1(r,a)$ is a monotonic \f,
because $\PD G_1 r >0$, and for this it has an inverse \f\ $G(r,a)$ \st
\beq2.39a
G(G_1(r,a),a)=r.
\e
Moreover, the parametrization of a characteristic line is not very convenient
for our purposes. Thus we should go to an old parameter~$t$. One easily gets
\beq2.40
r(t)=G\biggl(\frac{p_0}m\,\root3\of{\frac{d^2}{2k}}\,t+f,\root3\of
{\frac{d^2k^2}2}\biggr).
\e
Afterwards we should substitute Eq.\ \er{2.40} to Eqs \er{2.30}--\er{2.34}.

Let us notice the following fact: $T_0(t,r,p_r,\th,p_\th,\vf,p_\vf)$, a~\so\
of a \hg\ \eq\ is \cn\ along a \ch\ line
\beq2.41
\PD{} t \,T_0\bigl(t,r(t),p_r(r(t)),\th(r(t)),p_\th(r(t)),\vf(r(t)),
p_\vf(r(t))\bigr)=0.
\e
The last statement is equivalent to Eq.~\er{2.26} and Eqs (\ref{2.27}.1--7).
Thus we can consider the following initial problem
\beq2.42
T_0(0,r,p_r,\th,p_\th,\vf,p_\vf)=\wt F(r,p_r,\th,p_\th,\vf,p_\vf)
\e
where $\wt F$ is an arbitrary \f\ of six real \va s of $C^1$ class.

The \so\ of Eq.~\er{2.26} reads
\beq2.43
T(t,r,p_r,\th,p_\th,\vf,p_\vf)=t+T_0(t,r,p_r,\th,p_\th,\vf,p_\vf)
\e
($t$ is a special \so\ to Eq.~\er{2.26}.

{\allowdisplaybreaks
The graph of the \so\ of a \hg\ version of
Eq.~\er{2.26}---$T_0(t,r,p_r,\th,p_\th,\vf,p_\vf)$ must be a union of \ch s.
In order to find a \so\ of our \eq\ we should find $r(0)$, $p_r(r(0))$,
$\th(r(0))$, $p_\th(r(0))$, $\vf(r(0))$, $p_\vf(r(0))$. In this way
\beq2.44
T(t,r,p_r,\th,p_\th,\vf,p_\vf)=t+
\wt F\bigl(r(0),p_r(r(0)),\th(r(0)),p_\th(r(0)),\vf(r(0)),p_\vf(r(0))\bigr).
\e
One gets
\bea2.45
r(0)&=G\biggl(\cG_1(r)-\frac{p_0}m
\,\sqrt{\frac{d^2k^2}2}\,,\root3\of{\frac{d^2k^2}2}\biggr)\\
p_r(r(0))&=p_r \exp\Bigl(k\Bigl(\frac 1r-\frac1{r(0)}\Bigr)
+\frac{d^2}2\,(r^2-r^2(0))\Bigr) \lb2.46 \\
\th(r(0))&=\am\Bg({\frac{2k}{p_0}\Bg({\cG_2(r(0))-
\cG_2(r)+F\Bg(\th \Bigm| \frac{c^2}{d^2})})
\Bigm| \frac{c^2}{d^2} }) \lb2.47 \\
p_\th(r(0))&=\sqrt{p_\th^2+c^2\Bg({
\sn^2\bgg({\frac{2k}{p_0}\,\cG_2(r)+e \Bigm|\frac{c^2}{d^2}})
-\sn^2\Bg({\frac{2k}{p_0}\,\cG_2(r(0))+e \Bigm|
\frac{c^2}{d^2}})})} \lb 2.48 \\
\vf(r(0))&=\vf + \frac dc\Bg({\frac{2k}{p_0}\,\cG_2(r(0))-\cG_2(r) })
+E\bgg({\am\Bg({\frac{2k}{p_0}\,\cG_2(r)+e\Bigm|\frac{c^2}{d^2} })
\Bigm|\frac{c^2}{d^2}}) \nn \\
&\hskip160pt {}-E\bgg({\am\Bg({\frac{2k}{p_0}\,\cG_2(r(0))+e\Bigm|\frac{c^2}{d^2} })
\Bigm|\frac{c^2}{d^2}}) \lb2.49
\e
where
\bgr2.50
c=p_\vf\\
p_\vf(r(0))=p_\vf \lb2.51 \\
\cG_i(x)=G_i\bgg(x,\root3\of{\frac{d^2k^2}2}), \q i=1,2. \lb 2.d
\e
$\wt F$ is an arbitrary \f\ of $C^1$ class,
\bgr2.52
p_0=p_r \exp\Bg(\frac kr+\frac{d^2r^2}2)\\
e=F\Bg(\th\Bigm|\frac{c^2}{d^2})-\frac{2k}{p_0}\,\cG_2(r) \lb2.53 \\
d^2=p_\th^2+p_\vf^2\sin^2\th. \lb2.54
\e

In Fig.~1 we give a 3D-plot of the \f\ $G_1(r,a)$
and in Fig.~2 we give a 3D-plot of the \f\ $G_2(r,a)$.

\centerline{\includegraphics[height=240bp]{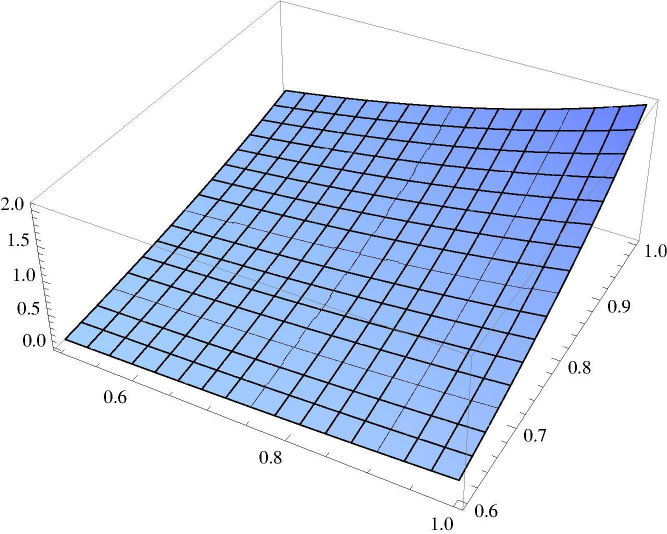}}
\nobreak \centerline{\small Fig.\ 1. \ 3D-plot of the \f\ $G_1(r,a)$ (see text)}

\centerline{\includegraphics[height=240bp]{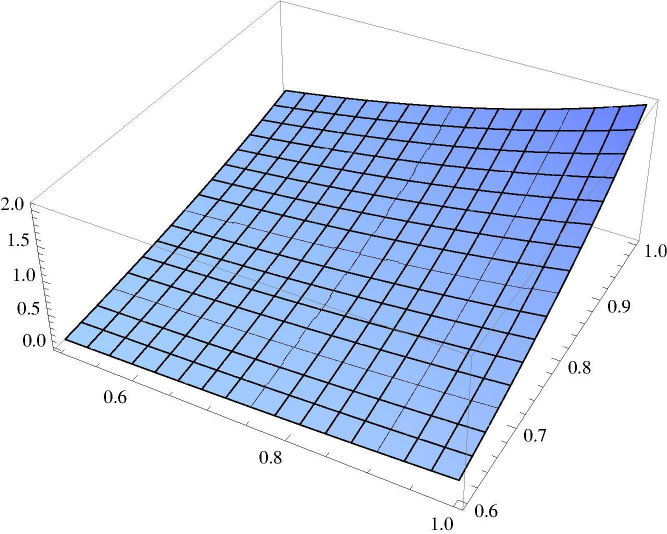}}
\centerline{\small Fig.\ 2. \ 3D-plot of the \f\ $G_2(r,a)$ (see text)}

\bigskip
We should substitute Eqs \er{2.50}, \er{2.52}, \er{2.54} into Eq.~\er{2.53}
getting
\beq2.55
e=F\Bg(\th \Bigm| \frac{p_\vf^2}{\bar p_{\vf\th}})
-\frac{2k}{p_r}\, G_2\Bg(r,\wt p_{\vf\th}^2)
\exp\Bg({-\Bg(\frac kr + \frac{r^2\bar p{}_{\vf\th}^2}2)}),
\e
where
\beq2.55d
\bar p_{\vf\th}^2=p_\th^2+p_\vf^2\sin^2\th, \q
\wt p_{\vf\th}^2 = \root3\of{\frac{k^2 \bar p_{\vf\th}^2}2}.
\e
Afterwards we substitute Eqs \er{2.50}, \er{2.52}, \er{2.54} and \er{2.55}
into Eqs \er{2.46}--\er{2.49}:
\bgr2.56
p_r(r(0))=p_r\exp\bgg({k\Bg(\frac1r -\frac1{r(0)})
+\frac{\bar p_{\vf\th}}2(r^2-r^2(0))})\\
\th(r(0))=\am\bgg({\frac{2k}{p_r}\exp\Bg({-\Bg({\frac kr
+\frac{r^2}2\,\bar p_{\vf\th}})}) \cdot G_2\bg({r(0),\wt p_{\vf\th}})
\Bigm|\frac{p_\vf^2}{\bar p_{\vf\th}}})\hskip50pt \nn \\
\hskip200pt {} -G_2\bg(r,\wt p_{\vf\th})
+F\Bg(\th\Bigm| \frac{p_\vf^2}{\bar p_{\vf\th}})\lb2.57 \\
p_\th(r(0))=\biggl[p_\th^2 + p_\vf^2\bgg({
\sin^2\th - \sn^2\bgg({\frac{2k}{p_r\exp(k/r+\bar p_{\vf\th}r^2/2)}\,
\Bg({G_2\bg({r(0),\wt p_{\vf\th}}) -G_2\bg({r,\wt p_{\vf\th}}) \nn \\
\hskip200pt{}+F\Bg(\th\Bigm|\frac{p_\vf^2}{\bar p_{\vf\th}})})
\biggm| \frac{p_\vf^2}{\bar p_{\vf\th}}})})\biggr]^{1/2} \lb2.58 \\
\vf(r(0))=\vf + \frac{\sqrt{\bar p_{\vf\th}}}{p_\vf}
\Biggl(\frac{2k}{p_r\exp(k/r+r^2\bar p_{\vf\th}/2)}
\bgg({G_2\bg({r(0),\wt p_{\vf\th}})-G_2\bg(r,\wt p_{\vf\th})})
+E\Bg(\th\Bigm| \frac{p_\vf^2}{\bar p_{\vf\th}})\hskip20pt \nn\\
\hskip20pt {}
-E\bgg({\am\Bg({\frac{2k}{p_r\exp(k/r+r^2\bar p_{\vf\th}/2)}
\Bg({G_2\bg({r(0),\wt p_{\vf\th}})
-G_2\bg(r,\wt p_{\vf\th})})\Bigm|\frac{p_\vf^2}
{\bar p_{\vf\th}}})\biggm|\frac{p_\vf^2}{\bar p_{\vf\th}}})\Biggr). \lb2.59 \\
p_\vf(r(0))=p_\vf \lb2.60
\e
and $r(0)$ is given by the formula \er{2.45}.

If we substitute Eqs \er{2.55d}--\er{2.60} and \er{2.45} into Eq.~\er{2.44},
we get a \so\ of an in\hg\ \eq\ \er{2.26} with the initial condition
\er{2.42}.

After all these substitutions we get eventually
\beq2.61
T(t,r,p_r,\th,p_\th,\vf,p_\vf)=t+\wt F(x_1,x_2,x_3,x_4,x_5,x_6)
\e
where $\wt F$ is a \f\ of $r,p_r,\th,p_\th,\vf,p_\vf$ via $x_i$,
$i=1,2,3,4,5,6$ and also a \f\ of~$t$ via $r(0)$ (see~Eq.~\er{2.45}).

Let us do the following substitution:
\bgr2.62
\vec L=\vec r\t \vec p\\
\vec p=p_r\vec e_r + p_\th \vec e_\th + p_\vf \vec e_\vf \lb2.63 \\
\vec r=r\vec e_r \lb2.64
\e
where $\vec e_r$, $\vec e_\th$ and $\vec e_\vf$ are tangent vectors to \cd\
lines in spherical \cd \ system
\bea2.65
\vec e_r&=\sin\th \cos\vf \,\vec e_x + \sin\th \sin\vf \,\vec e_y + \cos\th \,\vec
e_z \\
\vec e_\th&=\cos\th \cos\vf \,\vec e_x + \cos\vf \sin\vf \,\vec e_y
- \sin\th \,\vec e_z \lb2.66 \\
\vec e_\vf&=-\sin\vf \,\vec e_x + \cos\vf \,\vec e_y, \lb2.67
\e
$\vec e_x$, $\vec e_y$, $\vec e_z$ are versors in Cartesian \cd\ system.
In this way one gets
\bea2.68
L_x&=-r(p_\th \sin\vf + p_\vf \cos\th\cos\vf) \\
L_y&=r(p_\th\cos\vf - p_\vf \cos\th\sin\vf) \lb2.69 \\
L_z&=-rp_\vf \cos\th \lb2.70 \\
L^2&=p_\th^2 + p_\vf^2 \lb2.71
\e
and
\bea2.72
p_\th &= \frac1r (L_y\cos\vf - L_x\sin\vf) \\
p_\vf &= -\frac1{r\cos\th}( L_x\cos\vf + L_y\sin\vf). \lb2.73
\e
Substituting Eqs \er{2.72}--\er{2.73} into Eq.~\er{2.61} we can express $T$
in terms of $p_r$ and $L_x$ and~$L_y$.}

Let us consider a general Hamiltonian $H=H(q_i,p_i)$, $i=1,2,\dots,n$. Let us
come back to Eq.~\er{2.5} and write it according to Eq.~\er{2.3}. One gets
\beq2.79
\pd T t + \sum_{i=1}^n \Bigl(\pd T q_i \, \pd H p_i - \pd T p_i\, \pd H q_i
\Bigr)=1.
\e
Eq. \er{2.79} is a linear nonhomogeneous partial differential \eq\ of the first
order. According to general method of treatment of such an \eq\ we consider a
Cauchy initial problem
\beq2.80
T(q_i,p_i,0)=F(q_i,p_i), \q i=1,2,\dots,n,
\e
where $F$ is a \f\ of $2n$ variables of class $C^1$.

The problem can be solved in the following way. Let us consider the Cauchy
problem \er{2.80} for a homogeneous \eq
\beq2.81
\pd T t + \sum_{i=1}^n \Bigl(\pd T q_i \,\pd H p_i - \pd T p_i \,
\pd H q_i \Bigr)=0.
\e
The \so\ of the problem \er{2.80} for Eq.~\er{2.79} is given by the formula
\beq2.82
T(q_i,p_i,t)=t+\wt T(q_i,p_i,t),
\e
where $\wt T(q_i,p_i,t)$ is a \so\ of the problem \er{2.80} for Eq.~\er{2.81}.

Eq.~\er{2.81} can be solved by using a \ch\ method. Let us write the \eq\ for
a \ch\ curve for Eq.~\er{2.81}. One gets
\beq2.83
\bal
\PD t {\tau} &=1\\
\PD q_i {\tau} &=\pd H p_i \\
\PD p_i {\tau} &=-\pd H q_i ,
\eal
\e
$i=1,2,\dots,n$. $\tau$ is a parameter of a \ch.

The \so\ of Eq.~\er{2.81} is \cn\ along the \ch
\beq2.84
\PD {} {\tau} \wt T\bigl(t(\tau),q_i(\tau),p_i(\tau)\bigr)=0.
\e
The graph of a \so\ of Eq.~\er{2.80} must be a union of \ch s. If we choose a
curve $\Gamma$ in $(2n+1)$-dimensional space $\xi^0(\eta)$, $\xi^i(\eta)$,
$\xi^j(\eta)$, $i,j=1,2,\dots,n$, and $\eta$ is a parameter of the curve
we can drive at any point of the curve $\Gamma$ a \ch. In this way we get a \so\ in
terms of~$\tau$ and~$\eta$ parameters
\beq2.85
\wt T\bigl(t(\tau,\eta),q_i(\tau,\eta),p_i(\tau,\eta)\bigr),
\e
which is \cn\ along a \ch. Now it is necessary to satisfy the Cauchy initial
problem
\beq2.86
\wt T\bigl(0,q_i(\tau,\eta(\tau)),p_i(\tau,\eta(\tau))\bigr)
=F\bigl(q_i(\tau,\eta(\tau)),p_i(\tau,\eta(\tau))\bigr)
\e
where
\beq2.87
t(\tau,\eta(\tau))=0
\e
and $\eta(\tau)$ is a \so\ of Eq.~\er{2.87}.

In this way we get a \f\ $T(t,q_i,p_i)$ in a general case. It means a time
for any Hamiltonian.

Let us notice that a quantity $T$ is really a time. We can measure a~change
of any physical quantity \wrt $T$ as \wrt $t$. It is coming from Eq.~\er{2.2}.
One gets
\beq2.88
\PD{\gG} t = \PD {\gG} t \cdot \Bigl(\PD T t \Bigr)^{-1} = \PD {\wt\gG} T
\e
where
\beq2.89
\bga
\gG = \gG(q_i(t),p_i(t),t),\\
\wt\gG = \gG(q_i(T),p_i(T),T).
\ega
\e
$T$ is any solution of Eq.~\er{2.5} for our Hamiltonian system. In the case
of quantum mechanics because $T$ is a mechanical quantity (not a~parameter
as~$t$) it is quite easy to quantize it. We do it in the next section.

Let us notice the following fact. Any physical quantity $\gG_1$ which changes
\wrt time changes also \wrt a different physical quantity $\gG_2$, i.e.
\bgr2.90
\PD \gG_1 \gG_2 = \PD \wt\gG_1 \wt\gG_2 \\
\bal
\gG_k &= \gG_k(q_i(t),p_i(t),t), \q k=1,2,\\
\wt\gG_k &= \wt\gG_k(q_i(T),p_i(T),T), \q k=1,2,\ i=1,2,\dots,n.
\eal \lb2.91
\e
A notion of time is coming from this changing.

\section{Quantum \op\ of time}
Let us quantize $T(q,p,t)$ in the case of $n=1$, i.e.\ let us find an \op\
acting in a Hilbert space for
\beq3.1
T(q,p,t)=t+\wt F\Bigl(p\cos\o t+q\sqrt{Km}\sin\o t,-q\cos\o t+p\,\frac{\sin\o t}
{\sqrt{Km}}\Bigr).
\e
Now we consider $q$ and $p$ as \qm\ \op s acting in Hilbert space and
satisfying a commutation relation
\beq3.2
[\wh q,\wh p]=i\hbar,
\e
where $\wh q=q$, $\wh p=-i\hbar \frac d{dq}$.

In order to proceed this programme we suppose that $F(x,y)$ is an analytic
\f\ of~$x$ and~$y$, i.e.
\beq3.3
F(x,y)=\sum_{\dwa{n=0\\k=0}}^\iy a_{nk}x^ny^k.
\e
The usual procedure consists in replacing $q^np^k$ by powers of \op s $\wh q$
and~$\wh p$. There are three well known approaches, i.e.\ normal, antinormal
and Weyl (see Refs \cite{2n, 3n}).

In our case one gets
\bmlg
x^ny^k=\bigl(p\cos\o t+q\sqrt{Km}\sin\o t\bigr)^n
\cdot\Bigl(-q\cos\o t +p\,\frac{\sin\o t}{\sqrt{Km}}\Bigr)^k\hskip80pt\\
{}=\sum_{i=0}^n \binom ni p^i(\cos\o t)^i q^{n-i}(Km)^{(n-i)/2}(\sin\o t)^{n-i}\\
\hskip80pt {}\t \sum_{j=0}^k \binom kj(-1)^jq^j(\cos\o t)^j p^{k-j}(\sin\o t)^{k-j}
(Km)^{(j-k)/2}\\
{}=\sum_{i=0}^n \sum_{j=0}^k \binom ni \binom kj p^i(\cos\o t)^i q^{n-i}
(Km)^{(n-i)/2}(\sin\o t)^{n-i}\hskip80pt\\
\hskip80pt{}\t (-1)^jq^j(\cos\o t)^jp^{k-j}(\sin\o t)^{k-j}
(Km)^{(j-k)/2}\\
{}=\sum_{i=0}^n \sum_{j=0}^k (-1)^j \binom ni \binom kj p^{k+i-j}q^{n+j-i}
(\cos\o t)^{i+j} (\sin\o t)^{n+k-i-j}(Km)^{(n-k-i+j)/2}.
\e
Thus one gets
\bml3.4
\ov F(q,p)=\sum_{\dwa{n=0\\k=0}}^\iy a_{nk}\\
{}\t\sum_{i=0}^n \sum_{j=0}^k (-1)^j
\binom ni \binom kj p^{k+i-j}q^{n+j-i}
(\cos\o t)^{i+j} (\sin\o t)^{n+k-i-j}(Km)^{(n-k-i+j)/2}.
\e
In this way we get
\bml3.5
\wh T=t\\
{}+\sum_{\dwa{n=0\\k=0}}^\iy a_{nk}
\sum_{i=0}^n \sum_{j=0}^k (-1)^j
\binom ni \binom kj (Km)^{(n-k-i+j)/2}
(\cos\o t)^{i+j} (\sin\o t)^{n+k-i-j}\wh p{}^{k+i-j}\wh q{}^{n+j-i}
\e
where we choose the simplest possibility to replace $p^{l_1}q^{l_2}$ by the
\op s $\wh p{}^{l_1}\wh q{}^{l_2}$. We can replace $p^{l_1}q^{l_2}$ by
$\wh q{}^{l_2}\wh p{}^{l_1}$ or $\frac12\bg(\wh p{}^{l_1}\wh q{}^{l_2}
+\wh q{}^{l_2}\wh p{}^{l_1})$.

Let us consider $T$ in a different representation, i.e.\ \er{2.22} for $n=1$.
One gets
\beq 3.6n
\wt T(t,a,a^*)=t+\wt F\Bigl(-\sqrt{\frac{m\o\hbar}2}\,(ae^{i\o t}-a^*e^{-i\o t})
-\sqrt{\frac{\hbar}{2m\o}}\,(ae^{i\o t}+a^* e^{-i\o t})\Bigr).
\e
Using the expansion \er{3.3} one gets
\bml3.7n
T(t,a,a^*)=t\\
{}+\sum_{\dwa{n=0\\k=0}}^\iy a_{nk}(-1)^k
\sqrt{\frac{\hbar^{n+k}\o^{n-k}m^{n-k}}{2^{n+k}}}
\sum_{l=0}^n \sum_{j=0}^k \binom nl \binom kj (-1)^l
a^{l+j}a^{*(n+k-l-j)}e^{i(2l+2j-n-k)\o t}.
\e

Let us quantize $T$ in a different way substituting $a$ and~$a^*$ by
annihilation and creation \op s \st
\bgr3.8n
[\wh a,\wh a{}^+]=1 \\
a\mt \wh a, \q a^*\mt \wt a^+. \lb 3.9n
\e
One gets
\bgr3.10
\bal \wh a&=\sqrt{\frac{m\o}{2\hbar}}\,\Bigl(\wh q+\frac i{m\o}\,\wh p\Bigr)\\
\wh a{}^+&=\sqrt{\frac{m\o}{2\hbar}}\,\Bigl(\wh q-\frac i{m\o}\,\wh p\Bigr)
\eal \\
\bal \wh q&=\sqrt{\frac{\hbar}{2m\o}}\,(\wh a+\wh a^+)\\
\wh p&=i\sqrt{\frac{m\o\hbar}{2}}\,(\wh a-\wh a^+)
\eal \lb3.11 \\
\bal \wh a{}^+\bigl|n\bigr\rangle&=\sqrt{n+1}\,\bigl|n+1\bigr\rangle\\
\wh a\bigl|n\bigr\rangle&=\sqrt n\,\bigl|n-1\bigr\rangle
\eal \lb3.12
\e
where $\bigl|n\bigr\rangle$ represents a canonical base of a harmonic
oscillator, $\bigl|0\bigr\rangle$ being a vacuum state
\beq3.13
\wh a\bigl|0\bigr\rangle=0.
\e

The quantization procedure consists here in replacing $a^{l_1}a^{*l_2}$ by
a~product of \op s $\wh a,\wh a{}^+$. The simplest choice is to take $\wh
a{}^{l_1} \wh a{}^{+l_2}$. Moreover, there are different important
possibilities. We can substitute $a^{l_1}a^{*l_2}$ by $\frac12\bg(\wh
a{}^{l_1} \wh a{}^{+l_2}+ \wh a{}^{+l_2}\wh a{}^{l_1})$ or $\wh a{}^{+l_2}\wh
a{}^{l_1}$.

We can also use Weyl--Wigner transform to get a time \op
\beq3.14
\wh T = \frac1{2\pi} \int_{-\iy}^{+\iy} T(t,q,p)
\exp\bigl(i\bigl(a(\wh q-q)+b(\wh p-p)\bigr)\bigr)\,dq\,dp\,da\,db
\e
where $\wh q$ and $\wh p$ are generators of a Heisenberg algebra, and
$\hbar=1$.

One can get the expectation value of an \op~$\wh T$ as a trace of a product
of a density matrix $\wh \rho$ and~$\wh T$.

One can rewrite Eq.~\er{3.14} in a more convenient form defining an \op
\beq3.15
\wh B(q,p)=\frac1{2\pi\hbar} \mathop{\intop\hskip-8pt\intop}_{-\iy\kern8pt}^{\kern8pt+\iy}
da\,db\,\exp\Bigl(\frac i\hbar \bigl(a(\wh q-q)+b(\wh p-p)\bigr)\Bigr)
\e
and
\beq3.16
\wh T=\frac1{2\pi\hbar} \mathop{\intop\hskip-8pt\intop}_{-\iy\kern8pt}^{\kern8pt+\iy}
T(t,q,p)\wh B(q,p)\,dq\,dp.
\e
Moreover,
\beq3.17
\frac1{2\pi\hbar} \mathop{\intop\hskip-8pt\intop}_{-\iy\kern8pt}^{\kern8pt+\iy}
\wh B(q,p)\,dq\,dp={\rm Id}.
\e
Id is the identity \op\ in a Hilbert space.

In this way one gets
\beq3.18
\wh T=t+\frac1{2\pi\hbar}\mathop{\intop\hskip-8pt\intop}_{-\iy\kern8pt}^{\kern8pt+\iy}
\wt F\Bigl(p\cos\o t+q\sqrt{km}\sin\o t,
-q\cos\o t+p\,\frac{\sin\o t}{\sqrt{km}}\Bigr)\wh B(q,p)\,dq\,dp
\e
and we do not suppose $F$ to be an analytic \f.

The expectation value of $\wh T$ is given by
\beq3.19
t+\frac1{2\pi\hbar}\mathop{\intop\hskip-8pt\intop}_{-\iy\kern8pt}^{\kern8pt+\iy}
\wt F\Bigl(p\cos\o t+q\sqrt{km}\sin\o t,
-q\cos\o t+p\,\frac{\sin\o t}{\sqrt{km}}\Bigr){\rm Tr}\bigl(\wh \rho \wh B(q,p)
\bigr)\,dq\,dp.
\e

How to quantize a time of Coulomb or Newton interaction, i.e.\ Eq.~\er{2.44}?
The best way is to consider an angular momentum of a system, i.e.\
$p_\vf$~and~$\vf$ in such a way that
$$
\wh p_\vf=-i\hbar\,\pd{} \vf , \q \wt \vf=\vf, \q [\vf,\wh p_\vf]=i\hbar.
$$
Afterwards to put $\wh p_\vf$ and $\vf$ into the \f\ \er{2.44}. Moreover,
this is highly hard task for $p_\vf$ and $\vf$~enters many places in the \so\
of Eq.~\er{2.26}, i.e.\ Eq.~\er{2.44}.

We can apply Eq.\ \er{3.14} and also Eq.\ \er{3.16} in such a way that $q=\vf$
and $p=p_\vf$. We get
\beq3.20
\wh T=t+\frac1{2\pi\hbar}\int_{-\iy}^{+\iy}\wt F(r,p_r,\th,p_\th,\vf,p_\vf)
\wh B(\vf,p_\vf)\,d\vf\,dp_\vf
\e
and the expectation value is equal to
\beq3.21
t+\frac1{2\pi\hbar}\int_{-\iy}^{+\iy}\wt F(r,p_r,\th,p_\th,\vf,p_\vf)
\Tr\bigl(\wh\rho \wh B(\vf,p_\vf)\bigr)\,d\vf\,dp_\vf.
\e
According to general formalism we define a \f\ on a phase space
\beq3.22
T_Q(t,q,p)=\Tr\bigl(\wh B(q,p)\cdot \wh T\bigr)=t+F_Q(q,p)
\e
where $F_Q(q,p)$ is a \f\ on a phase space
\bml3.23
F_Q(q,p)=\frac1{2\pi\hbar}\int_{-\iy}^{+\iy}
\wt F\Bigl(\bar p\cos\o t+\bar q\sqrt{km}\sin\o t,
-\bar q\cos\o t+\frac{\bar p\sin\o t}{\sqrt{km}}\Bigr)\\
{}\t\Tr\bigl(\wh B(q,p)\wh B(\bar p,\bar q)\bigr)\,dq\,dp
=\wt F\Bigl(p\cos\o t+q\sqrt{km}\sin\o t, -q\cos\o t
+\frac{p\sin\o t}{\sqrt{km}}\Bigr).
\e

How to quantize a time \f\ in a general case for Coulomb--Newton interaction,
i.e.\ Eq.~\er{2.61}? We can look for some generalization of canonical
quantization procedure in curvilinear \cd s (spherical \cd s) or we come back
to Cartesian \cd s and proceed quantization. We
choose the second possibility. Thus we write
\def\nrw#1{\hfill (\theequation\rm#1)}
\refstepcounter{equation} \label{3.24}
$$\dsl{
\li r=\sqrt{x^2+y^2+z^2} \nrw a\cr
\li \th=\arccos\Bigl(\frac z{\sqrt{x^2+y^2+z^2}}\Bigr) \nrw b\cr
\li \vf=\arctan\Bigl(\frac yx\Bigr). \nrw c\cr}
$$

For momentum one gets
\refstepcounter{equation} \label{3.25}
$$\dsl{
\li p_r=\frac{p_x+p_y+p_z}{\sqrt{x^2+y^2+z^2}} \nrw a\cr
\li p_\th=\frac{xzp_x+yzp_y-(x^2+y^2)p_z}{\sqrt{x^2+y^2}\sqrt{x^2+y^2+z^2}}
\nrw b \cr
\li p_\vf=\frac{-yp_x+xp_y}{\sqrt{x^2+y^2}}\,. \nrw c\cr}
$$
Using Eqs (\ref{3.24}a--c) and Eqs (\ref{3.25}a--c) one finds
\beq3.26
\wt T(t,x,p_x,y,p_y,z,p_z)=t+\wt F(x_1,x_2,x_3,x_4,x_5,x_6)
\e
where now $x_i$, $i=1,2,3,4,5,6$, are \f s of $x,y,z,p_x,p_y,p_z$ and~$t$.

Now we construct the \op\ $\wh B(x,y,z,p_x,p_y,p_z)$.
\bml3.27
\wh B(x,y,z,p_x,p_y,p_z)=\frac1{(2\pi\hbar)^3} \int_{\R^6}
\,da_x\,db_x\,da_y\,db_y\,da_z\,db_z\,\exp\Bigl(\frac i\hbar\bigl(
a_x(\wh x-x)+b_x(\wh p_x-p_x)\\
{}+a_y(\wh y-y)+b_y(\wh p_y-p_y)+a_z(\wh z-z)
+b_z(\wh p_z-p_z)\bigr)\Bigr).
\e
Using \er{3.27} we construct a quantum time \op
\beq3.28
\wt T=t+\frac1{(2\pi\hbar)^3} \int_{\R^6}
T(t,x,p_x,y,p_y,z,p_z)\wt B(x,p_x,y,p_y,z,p_z)\,dx\,dp_x\,dy\,dp_y\,dz\,dp_z
\e
or
\beq3.29
\wh T=t+\frac1{(2\pi\hbar)^3}\int_{\R^6}\wt F(x_1,x_2,x_3,x_4,x_5,x_6)
\wt B(x,p_x,y,p_y,z,p_z)\,dx\,dp_x\,dy\,dp_y\,dz\,dp_z.
\e
We can also define a \f\ on a phase-space
\beq3.31
\bga
T_Q(t,\vec r,\vec p)=\Tr(\wh B(\vec r,\vec p)\cdot \wh T)
=t+\Tr(\wh  B(\vec r,\vec p)\wh T_0),\\
\vec r=(x,y,z), \q \vec p=(p_x,p_y,p_z).
\ega
\e
We have of course canonical commutation relations
\beq3.32
[\wh x,\wh p_x]=i\hbar, \q [\wh y,\wh p_y]=i\hbar, \q [\wh z,\wh p_z]=i\hbar,
\e
and $\wh p_x=-i\hbar\pd{} x $, etc.

Let us come back to substitution \er{2.72}--\er{2.73} in Eq.~\er{2.61}. In
order to \qz e such a quantity we proceed in the following way. Let us
substitute into it
\bea3.33
\wh L_x&=i\hbar\Bigl(\sin\vf\,\pd{} \th{} +\cot\th \cos\vf\, \pd{} \vf{} \Bigr)\\
\wh L_y&=i\hbar\Bigl(-\cos\vf\,\pd{} \th{} +\cot\th \sin\vf\, \pd{} \vf{} \Bigr)
\lb 3.34
\e
and let us define $\wh T$ via formula \er{3.29}.

Now $F(x_1,x_2,x_3,x_4,x_5,x_6)$ is an \op\ with the substitution
\bea3.35
\wh p_\th&=\frac1r\bigl(\wh L_y \cos\vf - \wh L_x\sin\vf\bigr)\\
\wh p_\vf&=-\frac1{r\cos\th}\bigl(\wh L_x\cos\vf + \wh L_y\sin\vf\bigr)\lb3.36
\e
Moreover, we should still \qz e $r$ and $p_r$, i.e.\ $x,y,z$ and
$p_x,p_y,p_z$ via formulas (\ref{3.24}a) and (\ref{3.25}a) substituted to
Eq.~\er{2.61}. This can be achieved by Eq.~\er{3.29}. There is no guarantee
that all those procedures of \qz ation are equivalent. This shows us that we
get a very rich structure for an investigation.

The \op\ $\wh T$ can be considered as \qm\ \mc\ time. Taking an expectation value
at any state we get \qm\ fluctuations of time around~$t$, i.e.
\bml3.6
\<\vf\mid\wh T\mid\vf>=t+\sum_{\dwa{n=0\\k=0}}^\iy a_{nk}
\sum_{i=0}^n \sum_{j=0}^k (-1)^j\\
{}\t \binom ni \binom kj (Km)^{(n-k-i+j)/2}
(\cos\o t)^{i+j} (\sin\o t)^{n+k-i-j}\<\vf\mid \wh p{}^{k+i-j}\wh q{}^{n+j-i}
\mid\vf>.
\e
For a matrix $\rho$ we write
\bmlg
\Tr(\wh\rho\wh T)=T=t+\sum_{\dwa{n=0\\k=0}}^\iy a_{nk}
\sum_{i=0}^n \sum_{j=0}^k (-1)^j\\
{}\t \binom ni \binom kj (Km)^{(n-k-i+j)/2}
(\cos\o t)^{i+j} (\sin\o t)^{n+k-i-j}\Tr(\rho
\wh p{}^{k+i-j}\wh q{}^{n+j-i}).
\e
We can try also to diagonalise the \op\ $\wh T$ in a Hilbert space.
Moreover, it seems it has a continuous spectrum. It seems that we should take
the following order of $p$ and~$q$ \op s, i.e.
\beq3.7
p^{k+i-j}q^{n+j-i}\to \frac12\bigl(\wh p{}^{k+i-j}\wh q{}^{n+j-i}
+\wh q{}^{n+j-i}\wh p{}^{k+i-j}\bigr).
\e
In this way $\wh T$ is an Hermitian \op\ with a continuous spectrum.

In the case of a free motion one gets for a time \f
\beq3.8
T=t +\sum_{\dwa{n=0\\k=0}}^\iy a_{nk}\sum_{i=0}^n \sum_{j=0}^k (-1)^j
\binom ni \binom kj p^{k+i-j}q^{n+j-i}t^{n+i}m^{-(n+i)}.
\e
The time \op\ looks like
\beq3.9
\wh T=t +\sum_{\dwa{n=0\\k=0}}^\iy a_{nk}\sum_{i=0}^n \sum_{j=0}^k (-1)^j
\binom ni \binom kj \wh p{}^{k+i-j}\wh q{}^{n+j-i}t^{n+i}m^{-(n+i)}
\e
or we can do an exchange \er{3.7}.

Let us consider an expectation value at $\vf=\bigl|p\bigr\rangle$ for
$\wh T$ of a \ho\ in $\wh a{}^+,\wh a$ representation:
\bml5.10
\wh T(t,\wh a,\wh a{}^+) = t\\
{}+\sum_{\dwa{n=0\\k=0}}^\iy a_{nk} (-1)^k
\sqrt{\frac{\hbar^{n+k}\o^{n-k}m^{n-k}}{2^{n+k}}}
\sum_{l=0}^n \sum_{j=0}^k \binom nl \binom kj (-1)^l \wh a{}^{l+j}
\wh a{}^{+(n+k-l-j)}e^{i(2l+2j-n-k)\o t}.
\e
One gets
\bml5.11
\<p\mid \wh T(t,\wh a,\wh a{}^+)\mid p>= t+a_{00}\\
{}+\sum_{r=1}^\iy \sum_{n=1}^\iy
a_{n,2r-n}(-1)^n \frac{\hbar^r}{2^r} (\o m)^{n-r}
\sum_{l=0}\binom nl \binom{2r-n}{r-l} (-1)^l \prod_{\dwa{f=1\\g=1}}^r
\sqrt{(p+f)(p+r+g)}\\
{} + \sum_{r=1}^\iy \frac{2^r\hbar^r}{(\o m)^r}\,a_{02r}
\prod_{\dwa{f=1\\g=1}}^r \sqrt{(p+f)(p+r+g)}.
\e
In this way we get interesting results for a special type of \qz ation
of~$T$. The series \er{5.11} is divergent.

We can use also coherent states representations $\bigl|\a\bigr\rangle$,
$\a\in\mathbb C$,
$$
\bga
\wh a\bigl|\a\bigr\rangle=\a\bigl|\a\bigr\rangle\\
\hbox{or }\bigl|\a\bigr\rangle=e^{-|\a|^2/2} \sum_{n=0}^\infty
\frac{\a^n}{\sqrt{n!}} \bigl|n\bigr\rangle = e^{\a\hat a{}^+ -\a^*\hat a}
\bigl|0\bigr\rangle = e^{-|\a|^2/2}e^{\a\hat a{}^+}\bigl|0\bigr\rangle
=e^{-|\a|^2/2}\sum_{n=0}^\infty \frac{\a^n}{n!}\,(\wh a{}^+)^n
\bigl|0\bigr\rangle.
\ega
$$
In this way for a \ho
$$
\<\a \bigm| T(t,a\wh a,\wh a{}^+) \bigm| \a> =
t+a_{00}+\sum_{\dwa{n=1\\k=1}}^\infty a_{nk}(-1)^k
\sqrt{\frac{\hbar^{n+k}\o^{n-k} m^{n-k}}{2^{n+k}}}\,
(a^*)^{n+k-l-j}\a^{l+j}e^{i(2l+2j-n-k)\o t}.
$$
Moreover, in this case we use Eq.~\ref{5.10} with an exchange
$\wh a{}^{l+j}\wh a{}^{+(n+k-l-j)} \to \wh a{}^{+(n+k-l-j)}\wh a{}^{l+j}$.

Moreover, we can cut it taking
$r=r_{\max}$, $a_{n,2r-n}=a_{02r}=0$ for $r>r_{\max}$.
Moreover, starting from Eqs
\er{3.22}--\er{3.23} an expectation value is given by the formula
\beq5.24
\int_{-\iy}^{+\iy} dq\,dp\,W(q,p)T_Q(t,q,p)=t+\int_{-\iy}^{+\iy} dq\,dp\,
W(q,p)F_Q(q,p)
\e
where $W(q,p)$ is a Wigner quasiprobability \f
\beq5.25
W(p,q)=\frac1{\pi\hbar} \int_{-\iy}^{+\iy} \psi^*(q+\ov q)\psi(q-\ov q)
\exp\Bigr(\frac{2ip\ov q}{\hbar}\Bigr)\,d\ov q
\e
or in the case of a density matrix $\wh \rho$ (mixed states)
\beq5.26
W(q,p)=\frac1{\pi\hbar}\int_{-\iy}^{+\iy} \<q+\ov q\mid \wh \rho \mid q-\ov q>
\exp\Bigl(-\frac{2ip\ov q}{\hbar}\Bigr)\,d\ov q
\e
where $\<x \mid \psi>=\psi(x)$ (a standard notation).

Moreover, if we use two different time \f s $T_1$ and $T_2$ and if we
consider a product of two time \op s $\wh T_1$ and~$\wh T_2$, we can use a
$\ast$-product
\beq5.27
T_1(t,q,p)\ast T_2(t,q,p) = \Tr\bigl(\wh B(q,p)\wh T_1\wh T_2\bigr).
\e
In order to get third \f\ $T_3(t,q,p)$ corresponding to an \op~$\wh T_3$ we
write in a standard way
\beq5.28
T_1(t,q,p)\ast T_2(t,q,p) = T_3(t,q,p) =
T_1(t,q,p)\exp\Bigl(\frac{i\hbar}2\,\cP\Bigr) T_2(t,q,p),
\e
where
\beq5.29
\cP=\frac{\overleftarrow\partial}{\partial q}\,
\frac{\overrightarrow\partial}{\partial p}\,.
\e
All the ideas of \qz ation presented here can be found in Refs \cite{3n,4n,5n}.
In the case of a time \op\ \er{3.31} we can use a Wigner quasiprobability \f
\beq5.30
W(\vec r,\vec p)=\frac1{(\pi\hbar)^3} \int_{\R^3}\psi^*(\vec r+\vec r_0)
\exp\Bigl(\frac{2i\vec p\vec r_0}{\hbar}\Bigr)\,d^3\vec r_0
\e
or
\beq5.31
W(\vec r,\vec p)=\frac1{(\pi\hbar)^3} \int_{\R^3} \<\vec r+\vec r_0\mid
\wh\rho\mid\vec r-\vec r_0> \exp\Bigl(-\frac{2i\vec p\vec r_0}\hbar\Bigr)
d^3\vec r_0.
\e
We calculate the expectation value using the formula
\beq5.32
\int_{\R^6}d^3\vec r\,d^3\vec p\, W(\vec r,\vec p)T_Q(t,\vec r,\vec p)
= t+\int_{\R^6}d^3\vec r\,d^3\vec p\,W(\vec r,\vec p)T_{0Q}(t,\vec r,\vec p)
=\<T_Q(t,\vec r,\vec p)>.
\e

\section{Time as a \sp}
We get a ``time'' series
\beq4.1
T_t=\Tr(\wh \rho\wh T)
\e
or
\beq4.1a
T_t=\<\vf\mid \wh T\mid\vf>.
\e
In this way we can consider a time as a \sp\ with a particular realization
for every $\vf$ or~$\rho$ ($t$ means here a parameter).

One can define a variation
\beq4.2
\Var(T_t)=\Tr(\wh \rho\wh T{}^2)-\bigl(\Tr(\rho \wh T)\bigr)^2.
\e
In the case of a free motion one gets
\beq4.3
T_t=t+\frac12\sum_{\dwa{n=0\\k=0}}^\iy a_{nk}
\sum_{i=0}^k \sum_{j=0}^n \Bigl((-1)^j\binom ki\binom nj
\bigl(\Tr(\wh\rho\wh p{}^{k-i+j}\wh q{}^{n-j+i})
+\Tr(\wh\rho\wh q{}^{n-j+i}\wh p{}^{k-i+j}\Bigr).
\e
In the case of a harmonic oscillator one gets
\beq4.4
T_{t+nT}=T_t+nT
\e
for \er{4.1}--\er{4.1a}, where $\o=2\pi\nu=\frac{2\pi}T$, $n$ is an integer.

In this way a \qm\ fluctuation occurs for $t\in(0,T)$. In the case of a free
motion they are for all $t\in\R$.
Moreover, we get for a harmonic oscillator
\beq4.5
\Var(T_{t+nT})=\Var(T_t)
\e
where $n$ is an integer.

One can define also a ``covariance matrix''
\beq4.6
\Cov(T_t,T_s)=\bca
\Tr\bigl(\rho(\wh T_t-T_t)(\wh T_s-T_s)\bigr)\\
\<\vf\mid (\wh T_t-T_t)(\wh T_s-T_s)\mid \vf>.
\eca
\e
In the case of a harmonic oscillator one simply gets
\beq4.7
\Cov(T_{t+nT},T_{s+mT})=\Cov(T_t,T_s)
\e
where $n$ and $m$ are integers.

One can also define
\beq4.8
\Corr(T_t,T_s)=\frac{\Cov(T_t,T_s)}{\sqrt{\Var(T_t)\Var(T_s)}}
\e
with the same properties as \er{4.7} in the case of a \ho. In all the
formulas concerning a \ho\ we use the prescription \er{3.7n}. Quantum
fluctuations of~$\wh T$ strongly depend on the \f~$\ov F$.

Let us define Var for the \op\ $\wh T$ (see Eq.~\er{3.29}). One gets
\bml4.11
\Var(\wh T_t)=\biggl(
\int_{\R^6}d^3\vec r\,d^3\vec p\, W(\vec r,\vec p)T_Q(t,\vec r,\vec p)\biggr)^2
-\int_{\R^6}d^3\vec r\,d^3\vec p\, W(\vec r,\vec p)\Bigl(T_Q(t,\vec r,\vec p)
*T_Q(t,\vec r,\vec p)\Bigr)\\
{}=\biggl(
\int_{\R^6}d^3\vec r\,d^3\vec p\, W(\vec r,\vec p)T_{0Q}(t,\vec r,\vec p)\biggr)^2
-\int_{\R^6}d^3\vec r\,d^3\vec p\, W(\vec r,\vec p)\Bigl(T_{0Q}(t,\vec r,\vec p)
*T_{0Q}(t,\vec r,\vec p)\Bigr).
\e
For the covariance we get
\bml4.12
\Cov(\wh T_t,\wh T_s)=
\int_{\R^6}d^3\vec r\,d^3\vec p\, W(\vec r,\vec p)
\Bigl(T_Q(t,\vec r,\vec p)-\<\wh T_Q(t,\vec r,\vec p)>\Bigr)
\Bigl(T_Q(s,\vec r,\vec p)-\<\wh T_Q(s,\vec r,\vec p)>\Bigr)
\e
where
\beq4.13
\<\wh T_Q(t,\vec r,\vec p)>=\int_{\R^6}d^3\vec r\,d^3\vec p\,
W(\vec r,\vec p)T_Q(t,\vec r,\vec p)
\e
and eventually
\beq4.14
\Cov(\wh T_t,\wh T_s)=\int_{\R^6}d^3\vec r\,d^3\vec p\,
W(\vec r,\vec p)\bigl(T_{0Q}(t,\vec r,\vec p)-\<T_{0Q}(t,\vec r,\vec p)>
\bigr)\bigl(T_{0Q}(s,\vec r,\vec p)-\<T_{0Q}(s,\vec r,\vec p)>
\bigr)
\e
We can proceed some calculations of Var, Cov, Corr using the Moyal approach (Ref.~\cite{6n})
and Wigner \f\ not only in the case of a time \op\ for Coulomb or Newton
interactions but also for \ho s. This will be done elsewhere.

Let us consider a more general problem with Eq.~\er{2.3}, i.e.
\beq4.15
T(0,q_i,p_i)=F(q_i,p_i), \q i=1,2,\dots,n.
\e
Now the Hamiltonian of the system is arbitrary.

A \so\ of the initial problem \er{2.11} is given by a sum of a general \so\
of a \hg\ \eq\ and special \so\ of an in\hg\ problem. The special \so\ of an
in\hg\ \eq\ initial problem is simply~$t$ and the \so\ of a \hg\ \eq\ with
the condition \er{4.15} can be found by a method of \ch s.

Thus we have
\beq4.16
T(t,q_i,p_i)=t+T_0(t,q_i,p_i).
\e
It is easy to see that
\beq4.17
T(0,q_i,p_i)=F(q_i,p_i).
\e
In order to \qz e our time \f\ we use phase-space \qz ation procedure. In
this method we have still to do with \f s defined on phase-space, i.e.\
$T(t,q_i,p_i)$. Moreover, to find an expectation value we should use a Wigner
\f, i.e.\
\beq4.18
\<\wh T>=t+\int_{\R^{2n}}dq_1 \cdots dq_n\,dp_1 \cdots dp_n
T_0(t,q_i,p_i)W(q_i,p_i).
\e
Variation can also be calculated
\beq4.19
\Var(\wh T)=\<\wh T{}^2>-\<\wh T>^2=\<\wh T{}_0^2>-\<\wh T_0>^2,
\e
where
\bgr4.20
\<T_0>=\int_{\R^{2n}}dq_1 \cdots dq_n\,dp_1\cdots dp_n\,T_0(t,q_i,p_i)
W(q_i,p_i)\\
\<T_0^2>=\int_{\R^{2n}}dq_1 \cdots dq_n\,dp_1\cdots dp_n\,T_0^2(t,q_i,p_i)
W(q_i,p_i)
\lb4.21
\e
We can also write
\bml4.22
\Cov(T_t,T_s)=\int_{\R^{2n}}dq_1 \cdots dq_n\,dp_1\cdots dp_n\,W(q_i,p_i)
\bigl(T(t,q_i,p_i)-\<T_t>\bigr)\bigl(T(s,q_i,p_i)-\<T_s>\bigr)\\
{}=\int_{\R^{2n}}dq_1 \cdots dq_n\,dp_1\cdots dp_n\,W(q_i,p_i)
\bigl(T_0(t,q_i,p_i)-\<T_{0t}>\bigr)\bigl(T_0(s,q_i,p_i)-\<T_{0s}>\bigr).
\e

\U\qm\ time in our
approach has nothing to do with \qm\ time \op\ from Ref.~\cite4 and it has
no known commutation relation with a Hamiltonian.
It is interesting to find a relativistic (or even quantum gravity) analogue
of $T_t$.

Let us notice the following fact. In Ref.~\cite x the authors obtain a theory
of time coming from completely different approach, getting something similar
to our results, i.e.\ $t+\hbox{``fluctuations''}$. It means tic-tac of a clock
with fluctuations. They are using higher-dimensions approach. Maybe using
Kaluza--Klein Theory we can connect both approaches.

\section{Conclusions, prospects for further research and some remarks}
In the paper we consider time as a \mc\ quantity, i.e.
\beq5.1
T=T(t,p_i,q_i), \q i=1,2,\dots,n,
\e
where $q_i$ are positions and $p_i$ are momenta. In this way the \f~$T$
depends on the Hamilton \f\ of the system ($n$~is the number of degrees of
freedom of the system). We solve the \eq\ of motion for~$T$ in several cases,
i.e.\ for a Hamiltonian of \ho, free motion (a~kinetic energy only) and a
Newton or a Coulomb interaction of two material points (transformed to a
material point in Coulomb or Newton potential). Let us look on Eq.~\er{2.2}
and Eq.~\er{2.5} in more details. Eq.~\er{2.5} if we suppose that
$T=T(q_i,p_i)$ (no dependence on~$t$) tells us that $T$~is canonically
conjugated to~$H$---the Hamilton \f:
\beq5.2
\{T,H\}=1.
\e
This is similar for a position $q_i$ and conjugated momentum~$p_i$:
\beq5.3
\{p_j,q_i\}=-\d_{ij}.
\e
Thus if we want to \qz e a time in a canonical way we should use a
prescription
\beq5.4
\{\wh T,\wh H\}\mapsto \frac1{i\hbar}\,[T,H]
\e
(a Dirac \qz ation prescription).

We can also use a Moyal bracket which can be defined in two ways:
\beq5.5
\{T,H\}_M=\frac2\hbar\,T(q,p)\sin\Bg({\frac\hbar2\Bg(
\frac{\overleftarrow\partial}{\partial q}\,
\frac{\overrightarrow\partial}{\partial p}
-\frac{\overleftarrow\partial}{\partial p}\,
\frac{\overrightarrow\partial}{\partial q})})H(q,p)
\e
or
\beq5.6
\{T,H\}_M = \frac2{\hbar^3\pi^2}\int dp'\,dp''\,dq'\,dq''\,T(q+q',p+p')
H(q+q'',p+p'')\sin\Bg({\frac2\hbar(q'p''-q''p')})
\e
i.e.\ with a help of a differential \op\ of an infinite order (in this case
we suppose that $T$ and~$H$ are smooth \f s) or as an integral transform
($T$~and~$H$ are only Borel \f s).

In this case one gets
\beq5.7
\{T,H\}_M=\{T,H\} + O(\hbar^2).
\e
Thus $\{T,H\}_M=1+O(\hbar^2)$.

However we face a real problem: a Hamilton \op\ in Quantum Mechanics is
positively defined and a \so\ of commutation relation in $L^2(\R^n)$ is well
known and involves differentiation which is not positively defined.
For this there is not time \op\ in~QM with desirable properties, see Ref.~\cite9.

One can find the \so\ to this problem in QM in Ref.~\cite4. In Ref.~\cite{4}
one can find also a full discussion on a~time-\op\ in~QM and on uncertainty
relation for energy and time. One can find here also some references to the
time \op\ approaches which we do not follow. This is
nonrelativistic mechanics. In the relativistic case we have to do with the
following CCR (Canonical Commutation Relations)
\beq5.8
[\wh Q_{i\mu},\wh P_{j\nu}]=-i\hbar \d_{ij}\cdot \eta_{\mu\nu} \cdot {\rm Id}.
\e
Id is the identity \op\ in a Hilbert space. They are called RCCR
(Relativistic Canonical Commutation Relations). We should also suppose that
$\wh Q{}^+_\mu=\wh Q_\mu$, $\wh P{}^+_\mu=\wh P_\mu$.

Moreover, the theory must be Lorentz covariant. So we should have a unitary
representation of a Lorentz group in a Hilbert space $(U)$ (e.g.\
$L^2(\R^n)$) \st
\beq5.9
\bal
L_{\mu\nu}\wh Q{}^\mu &= U(L)^{-1}\wh Q_\nu U(L)\\
L_{\mu\nu}\wh P{}^\mu &= U(L)^{-1}\wh P_\nu U(L)\\
\wh Q{}^\mu &= \eta^{\mu\nu}\wh Q_\nu\\
\wh P{}^\mu &= \eta^{\mu\nu}\wh P_\nu
\eal
\e
$L=(L_{\mu\nu})$ is a matrix of Lorentz \tf. Thus we have three conditions:
RCCR, hermiticity of $\wh Q_\mu$ and $\wh P_\mu$ and covariance. These three
conditions cannot be simultaneously satisfied. This statement is known as a
Busch theorem (see Ref.~\cite{an}). $\eta_{\mu\nu}$~is a Minkowski tensor, $\eta_{\mu\nu}=(-,-,-,+)$.

We consider in the paper \qz ation of nonrelativistic time \f\ which
describes fluctuations of a parameter~$t$ during a motion. We use several
approaches to \qz e the \f. Moreover, it would be very interesting to examine
a motion of some subsystems with \qm\ and classical time \f. The subsystems
can interact. This problem will be investigated later.

Let us notice the following fact. One can consider \qm\ analogue of
Eq.~\er{2.2} as a Heisenberg \eq
\beq5a.10
1=\PD {\wh T} t =\pd {\wh T} t + \frac i{\hbar}\,[\wh H,\wh T].
\e
In this way we get an \eq\ for $\wh T$ as a differential \eq\ in a space of
\op s. Such \eq s are hard to manage. Thus our approach (to consider
classical \eq, to solve it and to \qz e the \so) seems to be more convenient
to manage. In order to define a relativistic time \op\ we should consider
$\wh Q_\mu$ for $\mu=4$ supposing that $\wh Q_\mu$ are Hermitian and
covariance relations are satisfied.

We can fulfil these two conditions. We need relativistic analogue of
Eq.~\er{2.2} and afterwards to \qz e a \so\ in a canonical way or to use
phase-space formulation to find expectation value. For different approach to
this problem see Ref.~\cite{13}. This approach (stochastic quantization) is
using a notion of a wave function as a function of two variables $p$~and~$q$
at once. In this way our ordinary wave functions in $p$~or~$q$ representation
are combined, \op s $\wh Q{}^i$, $\wh P{}^i$, $\wh Q{}^\mu$, $\wh P{}^\mu$ in
CCR and RCCR are more complex.

The relativistic analogue of Eq.~\er{2.2} can be easily written for a charged
particle in an electromagnetic field
\beq5a.27
H=c\sqrt{(\vec p - \tfrac qc \vec A)^2 + mc^2}+qA_4,
\e
where $q$ is a charge of a particle, $m$ its mass, and $\vec A$, $A_4$ are
vector and scalar potentials. Eq.~\er{2.2} now reads
\beq5a.28
1=\pd T t + \biggl(\pd T {\vec x} \cdot \pd H {\vec p} - \pd T {\vec p} \cdot
\pd H {\vec x} \biggr),
\e
or
\beq5a.29
c\biggl(\Bigl(\vec p-\frac qc\,\vec A\Bigr)^2 + mc^2\biggr)^{-3/2}
\Bigl(\vec p-\frac qc\,\vec A\Bigr)\biggl(\pd T {\vec x} -\pd T {\vec p} \,
\pd {\vec A} {\vec x} \biggr) - q\,\pd T {\vec p} \,\pd {A_4} {\vec x}
+\pd T t =1.
\e
Taking $\vec A=0$ and $qA_4=V$ one gets
\beq5a.30
\frac{c\vec p}{(\vec p{}^2+mc^2)^{3/2}}\,\pd T {\vec x} - \pd T {\vec p}\,
\pd V {\vec x} +\pd T t =1.
\e
This equation can be solved using characteristic method and $T$ should be
considered as $Q_4$, afterwards we \qz e it as before.

According to the general theory from Section~2 we consider a homogeneous \eq\
\beq5a.31
\pd T t +\frac{c\vec p}{(\vec p{}^2+mc^2)^{3/2}}\,\pd T {\vec x}
- \pd T {\vec p} \,\pd V {\vec x} =0
\e
and a \so\ of \er{5a.30} is given by
\beq5a.32
T(\vec x,\vec p,t)=t+\wt T(\vec x,\vec p,t)
\e
where $\wt T(\vec x,\vec p,t)$ is a \so\ of the problem \er{2.80} for
Eq.~\er{5a.31}. Eq.~\er{5a.31} can be solved using a characteristic method.
Let us write \eq s for a characteristic curve for Eq.~\er{5a.31}. One gets
\beq5a.33
\bal
\PD t {\tau} &=1\\
\PD {\vec x} {\tau} &=\frac{c\vec p}{(\vec p{}^2+mc^2)^{3/2}}\\
\PD {\vec p} {\tau} &=-\pd V {\vec x} \,.
\eal
\e
$\tau$ is a parameter of a characteristic. Now we proceed according to
\er{2.84}--\er{2.87} getting a \so\ to a homogeneous \eq.

There are some interesting monographs on time (see Refs \cite{10}, \cite{11},
\cite{12}).
In the mentioned monographs one can find a definition of a time according
to a definition by an abstraction (a~division by an equivalence relation which
is here a simultaneous relation of events in any established reference frame)
and also some considerations on a topology of a time including a topological
dimension of a~time ($0$-, $1$-, $2$-dimensional time) and a topological
branching of a~time. A~discrete time is of course $0$-dimensional. The
interesting point in our investigations is a problem of a discrete time. It
means the following problem
\beq5a.11
\wh T|\psi\rangle = T|\psi\rangle
\e
where $T$ is an eigenvalue of a time \op~$\wh T$. The solution does not exist
because $\wh T$ has a continuous spectrum in the Hilbert space $L^2(\R^n)$.
This is because of two reasons:

$1^\circ$ $\wh T=t+\wh T_0$ ($t$ is continuous real arbitrary number),

$2^\circ$ $\wh T_0$ has in general a continuous spectrum,

\noindent because it is a function of $\wh p_i$ and $\wh q_i$ which have continuous
spectra in $L^2(\R^n)$ Hilbert space. Moreover, we can consider Hilbert spaces
different from $L^2(\R^n)$, e.g.\ $L^2([a,b]^n)$, where $\wh p_i$ have discrete
spectra and $\wh q_i$ continuous. In $L^2([a,b]^n)$ all functions from the
domains of $\wh p_i$ and $\wh q_i$ have the same values on boundaries of a
compact set $[a,b]^n$, i.e.
\beq5a.12
f(q_1,\dots,q_i,a,q_{i+1},\dots,q_n)=f(q_1,\dots,q_i,b,q_{i+1},\dots,q_n).
\e
In particular for $n=1$
\beq5a.13
f(a)=f(b).
\e
$\wh p_i$ and $\wh q_i$ are Hermite \op s in $L^2([a,b]^n)$ Hilbert space and
the argument from Ref.~\cite4 does not work. $\wh q_i$~are bounded and
$\wh p_i$~are unbounded.

One gets
\beq5a.14
-i\hbar\,\frac{df}{dq}=\la f, \quad f(a)=f(b)
\e
and we find
\beq5a.15
f(q)=A\Bigl(\cos\Bigl(\frac{\la q}{\hbar}\Bigr) + i \sin\Bigl(\frac{\la q}
{\hbar}\Bigr)\Bigr),
\e
where
\beq5a.16
\la_k=\frac{2\pi\hbar}{(a-b)}, \quad k=0,\pm1,\pm2,\dots.
\e
In this way we solve an eigenvalue problem for a momentum \op.

Let us pose an eigenvalue problem for $T_0$ \op
\beq5a.17
\wh T_0 \lvert \psi_k\rangle=T_k \lvert \psi_k\rangle.
\e
This is possible if $\wh T_0$ is a function of~$\wh p$ only. In particular,
\beq5a.18
T=t+F(p)
\e
and
\beq5a.19
T_0=F(p)
\e
($F$ --- a real \f\ of one variable). One gets
\beq5a.20
\wh T=t+F(\wh p)=t+\wh T_0
\e
and
\beq5a.21
\wh T_0 \lvert \psi_k\rangle=F(\la_k)\lvert \psi_k\rangle
\e
where
\beq5a.22
\psi_k=A\Bigl(\cos\Bigl(\frac{\la_k q}{\hbar}\Bigr)
+ i \sin\Bigl(\frac{\la_k q}{\hbar}\Bigr)\Bigr), \quad
\la_k=\frac{2\pi k\hbar}{(a-b)}, \quad k=0,\pm1,\pm2,\dots,\quad
A=\frac1{\sqrt{|a-b|}}\,.
\e
Thus one gets:
\beq5a.23
T_k=F\Bigl(\frac{2\pi k\hbar}{a-b}\Bigr), \quad k=0,\pm1,\pm2,\dots,
\e
are eigenvalues of an \op~$\wh T_0$ and $\wh T_0$ has a discrete spectrum.
$\wh T_0$~is of course selfadjoint ($\wh T$~also).

In this way we can get several branches of time, similar as in Ref.~\cite{x}.
This gives us a possibility of topological branching points of time and some
quantum effects.

Time can be also discrete in a certain way
\beq5a.24
T=t+F(\la_k).
\e
The above simple example corresponds to $1$-dimensional free motion with
periodic boundary condition $(f(a)=f(b))$ and the corresponding Hamiltonian
\beq5a.25
H=\frac{p^2}{2m}
\e
has a discrete spectrum
\beq5a.26
E_k=\frac{2\pi^2\hbar^2k^2}{m(a-b)^2}\,.
\e

Let us consider the second example of a discrete time. This is a problem of
infinitely deep well potential in one dimension. One gets
\bgr 5a.34
V(x)=\bca
0& \hbox{for }x\in[-a,a]\\ +\infty &\hbox{otherwise,} \eca \nonumber \\
-\frac{\hbar^2}{2m}\cdot\PD {^2\psi} {x^2} =E\psi
\e
and
\beq5a.35
\psi(a)=\psi(-a)=0.
\e
Solving this problem we get
\bgr5a.36
\bca
\dfrac1{\sqrt a}\sin\o_k x & \hbox{for }x\in[-a,a],\\
0 &\hbox{otherwise,}
\eca\\
\o_k=\frac{k\pi}{a}\,, \qquad k=0,\pm1,\pm 2,\pm 3,\dots, \nonumber \\
E_k=\frac{\hbar^2k^2\pi^2}{2ma^2}\,. \lb5a.37
\e
A function of time $T$ now reads (this is a \so\ without explicit dependence
on~$x$, and even \wrt $p$)
\beq5a.38
T=t+F\Bigl(\frac{p^2}{2m}\Bigr)
\e
or
\beq5a.39
T(p,x,t)=t+F(H) \quad \hbox{for } x\in [-a,a].
\e
After \qz ation
\beq5a.40
\wh T=t+F(\wh H)
\e
which gives us
\beq5a.41
T_k=t+F(E_k).
\e
$F$ is a real arbitrary \f\ of one real variable.

Moreover, we refer to some quite recent papers (\cite{xx}, \cite{y}, \cite{z})
and an old work by W.~Pauli (Ref.~\cite{w}) on time-\op. J.~Kijowski
underlines that his time-\op\ constructed previously is self-adjoint
(Ref.~\cite{xx}). B.~Mielnik and G.~Torres-Vega pose a possibility to construct
such an \op\ using POV (Positive \U\op\ Valued) measure (see Ref.~\cite{z}).
J.~Kijowski criticizes (see Ref.~\cite{y}) W.~Pauli's (see Ref.~\cite{w})
arguments against quantum time-\op. Those arguments do not concern his
construction. Let us give some remarks on POV (a~semi-spectral measure) and
PV measures. It seems that POV measure can be reduced to PV measure (a~spectral
measure) via extension of a Hilbert space. The authors of Ref.~\cite{k} and
\cite{l} are not right and their arguments do not concern J.~Kijowski's
construction. J.~Kijowski's construction is unique. It follows from the
uniqueness of the \qz ation through Lie derivative which applies to any
observable linear in momenta.

E. Prugove\v cki uses POV (semispectral) measure in a place of PV (spectral)
measure. He uses also a vector generating representation as a wave \f\ (for
Galileo and Poincar\'e group) defined on a phase space.

The problem of a time in General Relativity is more complex. Someone can use
Loop Quantum Gravity approach (see Ref.~\cite{20}).

Moreover, in the case of Loop Quantum Cosmology (LQC) the situation is
simpler. In the case of unimodular gravity with a cosmological \cn\ (see
Ref.~\cite a) after reduction to a cosmological model one is able to
construct a 4-volume \op\ similar to Relativistic Time of Arrival Operator.
This \op\ can be considered as a cosmological time \op. It is self-adjoint
and has a discrete spectrum. The \op\ satisfies a commutation relation
\beq5.42
[\wh T_0,\wh\La] = 8\pi i l_p^2 \wh I = i\hbar \k \wh I,
\e
where $\wh\La$ is an \op\ of a cosmological \cn\ and $l_p$ is the Planck's
length. It means $\wh T_0$ is canonically conjugated to a cosmological \cn,
$\k=\frac{8\pi G_N}{c^4}$, $G_N$~is a Newton \cn, $c$~---~the velocity of
light, $\wh I$~---~the identity \op. According to Ref.~\cite a the
cosmological model under consideration is spatially flat, \hg\ and isotropic.
We have before \qz ation of $T_0$
\beq5.43
\dot T_0(t) = \{T_0,H\}.
\e
Moreover, in our approach
\beq5.44
\PD T t = \pd T t + \{T,H\}
\e
and
\beq5.45
T = t+T_0.
\e
After \qz ation one gets
\beq5.46
\wh T = t+\wh T_0
\e
with a spectrum
\beq5.47
T = t+\la_k.
\e
We can also play with a Moyal bracket
\beq5.48
\PD T t = \pd T t +\{T,H\}_M.
\e
This is a quantum analogue of Eq.~\er{5.44}.

Let us notice that we have to do with something called an ``arrow of time''.
It means a~time is directed. We have to do with three types of ``arrows of
time''. The first one is a thermodynamical arrow of time. A flow of time is
directed to states with higher entropy. The second ``arrow of time'' is a
cosmological arrow of time connected to expansion of the Universe (thus a
cosmological time is very important). The third arrow of time is connected
to PC-symmetry breaking in e.g.\ $K^0,\ov K{}^0$ decay
(Cabbibo--Kobayashi--Maskawa matrix) in elementary particle physics (see e.g.
Ref.~\cite c). If all processes in physics are invariant \wrt PCT-symmetry,
PC-symmetry breaking means that the particular process is not invariant \wrt
T-symmetry (time reversal symmetry). Thus there is not an invariance changing
$t$~to~$-t$, the same as in the case of increasing of entropy and expansion of
the Universe.

In Ref. \cite{aa} one considers a static space-time as a background of dynamics
of the Universe assuming that the Universe is determined by a unitarily
evolving wavepacket defined on a space-time. The model is considered in
$1+1$-dimensional case. Due to this formalism one can connect in some sense
two ``arrows of time'': cosmological (expansion of the Universe) and
thermodynamical (increasing of an entropy in the sense of Renyi entropy) via
spreading wavepackets in quantum mechanics. In our approach such a possibility
also exists.

In Refs \cite{bb}, \cite{cc} and references therein the problem of an existence of an operator
indicating the direction of time has been proved. It means an existence of an operator of an
``arrow of time'' in quantum mechanics. It has been proved that such an
operator exists and it is a self-adjoint operator. Such a self-adjoint
operator has the so-called Lyapunov property, i.e.\ monotonicity of the
expectation value irrespective of the initial state of the system. The
operator indicating the direction of time (i.e.\ an ``arrow of time'') can be
used to indicate the third arrow of time connecting to breaking of PC-symmetry
in elementary particle physics (a~time reversal symmetry breaking under an
assumption PCT invariance). Contemporary we have (see Ref.~\cite{CP}) a~direct
experimental proof of a time reversal symmetry breaking in $B^0$, $\ov B^0$
($B_s^0$, $\ov B_s^0$, $B_u^0$, $\ov B_u^0$, etc.) meson decays.

Moreover, we can define a WKB-time in semiclassical approximation in Quantum
Gravity in the following way. Let us follow C.~Kiefer (see Ref.~\cite{ab}):
\beq5.49
\pd{} t = \int d^3x\,G_{abcd}\,\frac{\d S_0}{\d h_{ab}}\,\frac{\d}{\d h_{cd}}\,,
\e
where
\bgr5.52
g_\m \,dx^\mu\,dx^\nu = \bigl(-N^2+\s_k \s^k\bigr)\,dt^2
+2\s_k\,dx^k \,dt + h_{ab}\,dx^a \,dx^b \\
S = MS_0+S_1+M^{-1}S_2 +\ldots \lb{5.50}
\e
is an action of gravity and matter,
\bgr5.51
M=\frac{c^2}{32\pi G_N}\\
G_{ijkl}=h_{ik}h_{jl} + h_{il}h_{jk} - h_{ij}h_{kl}. \lb{5.53}
\e
Eq.~\er{5.49} has been obtained from the Wheeler--DeWitt equation
\beq5.54
\bga
\biggl(-\frac{\hbar^2}{2M}\,G_{abcd}\,\frac{\d^2}{\d h_{ab}\d h_{cd}}
+MV + \cH_m\biggr)\Psi =0,\\
V=2c^2 \sqrt h\,(R-2\D),
\ega
\e
$\cH_m$ is a matter hamiltonian. In this case one gets
\beq5.55
1=\pd T t +\{T,\cH\}_M,
\e
i.e.
\beq5.55b
1= \int d^3x\,G_{abcd}\,\frac{\d S_0}{\d h_{ab}}\,\frac{\d T}{\d h_{cd}}
+\{T,\cH\}_M,
\e
where
\bgr5.56
\cH = \cH_G+\cH_m \\
\cH_G = \frac1{2\sqrt h}\,G_{ijkl}\pi^{ij}\pi^{kl} - \sqrt h\,R \lb{5.57}
\e
is a gravity hamiltonian.

In this case $h_{ab}$ are playing the role of~$q$ (coordinates) and
a momentum is defined by
\beq5.55a
\pi^{ab} = M\,\frac{\d S_0}{\d h_{ab}}\,.
\e
Moreover, our time~$T$ can be considered as a functional
\beq5.56a
T=T\bigl(\{h_{ab}\},\{\pi^{cd}\}\bigr).
\e
It is easy to see that $T$ is not an external parameter. It is really a
dynamical quantity.

Let us incorporate our time $T$ in the Ashtekar--Lewandowski formalism.
Let $\gd \Gamma,i,a,$ be a one-form $\rm so(3)$-valued connection (a~spin connection)
and $\gd K,j,a,$ an extrinsic curvature on~$M$ 3-manifold (a~space-like manifold
in canonical formalism for gravity, constant time $t$ manifold, see Ref.~\cite{20} and
also Refs \cite{k1,l1,m1}). $\gd\Gamma,j,a,$ and
$\gd K,j,a,$ are Ashtekar \va s ($i,a=1,2,3$).

Let $\gd A,a,j,$ be a Yang--Mills' $\rm SO(3)$ connection such that
\beq5.61
-\gamma \gd K,i,a, =\gd \Gamma,i,a, -\gd A,i,a, .
\e
The Ashtekar \va s are coming from a densitized inverse triad on~$M$ in such
a way that
\beq5.62a
gg^{ab} = \gd \wt E,a,i, \gd \wt E,b,i, \quad g=\det g_{ab},
\e
$g_{ab}$ is a metric tensor on~$M$, $\gd \wt E,a,i,$ is a densitized inverse
triad on~$M$ and
\beq5.63a
\partial_{[a}\gd e,i,a], = \gd\Gamma,i,[a, e_{b]}.
\e
$\gd e,i,a,$ is a triad on $M$.

The conjugate momentum to $\gd A,i,a,$ is $\gd P,a,j,$ and we have
($\gamma$~is an Immerzi--Bariaceri parameter)
\beq5.62
\{\gd A,i,a,(x),\gd P,b,j,(y)\} = \d^i_j \d^b_a \d(x,y),
\e
which is the only one nonvanishing Poisson bracket. The quantity $T$ is defined
($\cal H$ is a hamiltonian).
\beq5.63
1=\pd T t + \{T,{\cal H}\}, \q \PD T t = 1.
\e
We can write down the last equation using a Moyal bracket
\beq5.64
1 = \pd T t +\{T,{\cal H}\}_M.
\e

Eq.\ \er{5.64} can be rewritten
\beq5.56b
1=\pd T t + \sum_{i,a} \biggl(\Pd T A^i_a \,\Pd {\cal H} P^a_i - \pd T P^a_i
\,\pd {\cal H} A^i_a \biggr),\qquad \PD T t =1,
\e
where
\beq5.57b
{\cal H}= \int_M h\,d^3x.
\e
$\cal H$ is a general relativistic gravitational hamiltonian and $h$ its density on~$M$.
\beq5.58b
T= t+T_0\{A^i_a, P^a_i\}.
\e
We can also use a spinor version $A_a=-\frac12 A^i_a(x)\sigma_i$, $\wt E{}^a
=-i \wt E_i(x)\sigma_i$, $\sigma_i$ are Pauli matrices. The constant $\gamma$ can
be considered as a new constant of Nature connected to black holes entropy or
it can be considered for two values $\gamma=1$ or $\gamma=\sqrt{-1}$.

This can be developed in future papers.

At the end of the paper we quote a sentence by Roman Ingarden (see Ref.~\cite
b): ``{\it We are living under a destructive power of a time}''.

\section*{Acknowledgements}
I would like to thank Professor Bogdan Lesyng for inspiring discussions, in
particular, for pointing out that quantum and classical time, or time in the
interacting quantum and classical subsystems must be different mathematical
objects, as well as to thank him for providing such research conditions
in the BioExploratorium Centre of Excellence at the University of Warsaw
as well as in Bioinformatics Laboratory, IMDiK, Polish Academy of Sciences,
including access to Mathematica\TM~9\footnote{Mathematica\TM\ is the
registered mark of Wolfram Co.}, that this work could be effectively done.

\end{document}